\documentclass[letterpaper, 10 pt, conference]{ieeeconf}

\IEEEoverridecommandlockouts                              % This command is only
\usepackage{graphics} % for pdf, bitmapped graphics files
\usepackage{epsfig} % for postscript graphics files
\usepackage{amsmath} % assumes amsmath package installed
\usepackage{amssymb}  % assumes amsmath package installed

\usepackage{url}
\usepackage[ruled, vlined, linesnumbered]{algorithm2e}
\usepackage{verbatim} 
\usepackage{soul, color}
\usepackage{lmodern}
\usepackage{fancyhdr}
\usepackage[utf8]{inputenc}
\usepackage{fourier} 
\usepackage{array}
\usepackage{bm}
\usepackage{makecell}
\usepackage{amsmath}
\usepackage{mathtools}
\usepackage{bm}
\usepackage{amssymb}
\DeclareMathAlphabet\mathbfcal{OMS}{cmsy}{b}{n}

\SetNlSty{large}{}{:}

\makeatletter

\newcommand{\Rom}[1]{\expandafter\@slowromancap\romannumeral #1@}
\makeatother

\title{\LARGE \bf
Both orbital and spin torques originate from $\mathbf{r}\times\mathbf{g}$
}

\author{Frances Crimin and Stephen M. Barnett % <-this % stops a space 
\\ \\ \textit{Department of Physics and Astronomy, University of Glasgow,
Glasgow G12 8QQ, United Kingdom} \\
 \small{email: frances.crimin@glasgow.ac.uk}
}

\begin{document}

\maketitle
\thispagestyle{plain}
\pagestyle{plain}

%%%%%%%%%%%%%%%%%%%%%%%%%%%%%%%%%%%%%%%%%%%%%%%%%%%%%%%%%%%%%%%%%%%%%%%%%%%%%%%%
\begin{abstract}

Does the cross-product of the position and the electromagnetic momentum density, $\mathbf{g}$, include the optical spin momentum? We answer this long-standing question in the affirmative by evaluating, explicitly, the torque exerted on a particle by a beam of light carrying angular momentum. 
\\ 
\end{abstract}

%%%%%%%%%%%%%%%%%%%%%%%%%%%%%%%%%%%%%%%%%%%%%%%%%%%%%%%%%%%%%%%%%%%%%%%%%%%%%%%%
\section{Introduction}

It is now well-established that a beam of light can carry both spin and orbital angular momentum about the beam axis~\cite{Allen1992, OAMbook}. Experiments based on optical tweezers have demonstrated that a circuarly polarized beam will cause a trapped particle to rotate about its centre, and a Laguerre-Gaussian beam with an azimuthal phase dependence induces an orbital motion about the beam axis~\cite{ONeil2002}. Balancing these sources of spin and orbital angular momentum has shown, moreover, that the spin and orbital torques can have the same magnitude~\cite{Friese1996,Simpson1997} from which it may be inferred that the spin and orbital angular momenta carried by the light beam are quantised in units of the same fundamental quantity. 

The linear momentum density for a light beam in free space is $\mathbf{g}=\mathbf{E}\times\mathbf{H}/c^2$~\cite{Jackson1962}. From this we can construct the angular momentum density as $\mathbf{r}\times\mathbf{g}$. It has long been known how to separate this into spin and orbital parts~\cite{Darwin1932}, but this brings with it a paradox. This is the fact that for a plane wave, or indeed, for a phase-flat portion of a beam, $\mathbf{g}$ points in the direction of propagation and so $\mathbf{r}\times\mathbf{g}$ \textit{cannot} have a component in the direction of the beam~\cite{Khrapko2001,Simmons1970}. However, in response it has been shown that absorption of light from the beam induces a change in the angular momentum density that can be associated with the angular momentum gained by the particle ~\cite{Simmons1970, Allen2002,Zambrini2005}.  Indeed it had also previously been demonstrated by Feynman that a circularly polarized field \textit{does} transfer angular momentum to an absorbing particle~\cite{Feynman1965}.

The resolution of the ``plane-wave paradox" described above is in a sense unsatisfactory or, at least, incomplete in that it does not make quantitative contact with the torques produced by the beam. We address this issue here by a direct application of the local conservation of total angular momentum. This conservation law necessarily includes both the angular momentum density and the flux density of the angular momentum~\cite{Jackson1962,Barnett2002}. Indeed optically-induced torques can be measured indirectly by reconstructing the total angular momentum flux~\cite{Marsden,Strasser2022}.
The significance of the flux density is well illustrated by reference to a puzzle posed by Phillips~\cite{Phillips}, and it is worth pausing to present this.

\begin{figure}
\centering
\includegraphics[scale=0.3]{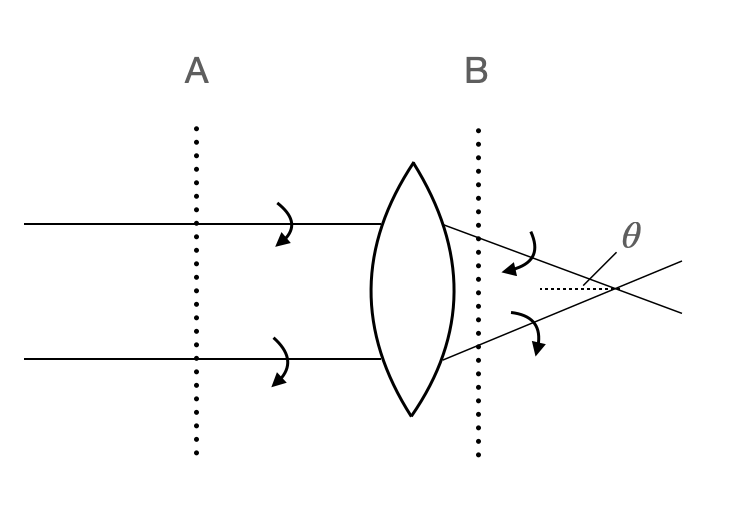}
\caption{A circularly polarized beam is brought to a focus through a lens. As a result, the component of the spin angular momentum in the direction of propagation appears to be reduced between planes A and B, but consideration of the \textit{flux} of the angular momentum resolves the problem of any ``missing" angular momentum.} 
\label{fig:phillipsparadox}
\end{figure}

Figure \ref{fig:phillipsparadox} represents the focusing of a circularly polarized beam. In the act of focusing the rays depicted are refracted towards a focal spot so that, for the upper-most ray depicted the component of the spin angular momentum in the direction of propagation is reduced by $\cos\theta$. It follows that the density of this component of the angular momentum after passing through the lens is less than it was before the lens. What has happened to the lost angular momentum? The answer is, nothing! The \textit{flux} of the angular momentum through plane A equals that through plane B because the velocity of light in the direction of the beam propagation for the upper-most ray is also reduced by a factor $\cos\theta$. This simple example serves to emphasise the importance of the angular-momentum flux in questions concerning the local conservation of the angular momentum. 

\section{Force calculation}
As a preliminary step, we consider first the force acting on a trapped particle. The local conservation law for the linear momentum follows directly from the microscopic Maxwell equations, 
\begin{equation}
\frac{\partial}{\partial t}\frac{1}{c^2}[\mathbf{E}\times\mathbf{H}]_i+ \nabla_jT_{ij}=-f_i, 
\end{equation}
where $T_{ij}$ is the electromagnetic momentum flux density
\begin{equation}\label{eq:momfluxdensity}
T_{ij}=\frac{1}{2}\delta_{ij}(\epsilon_0E^2+\mu_0H^2)-\epsilon_0E_iE_j-\mu_0H_iH_j,
\end{equation}
and we have employed the summation convention for repeated indices ranging over the three spatial cartesian coordinates. The source term is minus the Lorentz force density \cite{Lorentzbook}
\begin{equation}
\mathbf{f}=\rho\mathbf{E}+\mu_0\mathbf{J}\times\mathbf{H},
\end{equation}
where $\rho$ and $\mathbf{J}$ are the full, microscopic charge and current densities. We aim to determine the force, and from this the torque, acting on a small piece of dielectric held in an optical tweezer, as in the experiments mentioned above. To this end we introduce the polarization field $\mathbf{P}$, defined by the relations $\boldsymbol{\nabla}\cdot\mathbf{P}=-\rho$ and $\dot{\mathbf{P}}=-\mathbf{J}$, so that our force density becomes 
\begin{equation}
\mathbf{f}=-(\boldsymbol{\nabla}\cdot\mathbf{P})\mathbf{E}+\mu_0(\dot{\mathbf{P}}\times\mathbf{H}).
\end{equation}
This is not the most appropriate form of the force density for our purposes, so we rewrite it using Maxwell's equations, in the form~\cite{Barnett2006a}
\begin{equation}
f_i=P_j\nabla_iE_j-\nabla_j(P_jE_i)+\mu_0\frac{\partial}{\partial t}[\mathbf{P}\times\mathbf{H}]_i.
\end{equation}
At this point, our calculation is still exact, but we now specialize to the experimentally relevant situation in which the applied optical field is monochromatic and of sufficiently high frequency to allow cycle averaging, by discarding contributions oscillating at twice the optical frequency. To this end we write $E_i=\Re[\mathcal{E}_i\exp(-i\omega t)]$ and $P_i=\Re[\mathcal{P}_i\exp(-i\omega t)]$, so that our cycle-averaged force density becomes
\begin{equation}\label{eq:forcedensity}
f_i=\frac{1}{4}\left\{\mathcal{P}_j\nabla_i\mathcal{E}_j^*+\mathcal{P}_j^*\nabla_i\mathcal{E}_j-\nabla_i(\mathcal{P}_j\mathcal{E}_i^*+\mathcal{P}_j^*\mathcal{E}_i)\right\}.
\end{equation}
We observe forces rather than force densities (which are not unique \cite{Barnett2006a}), and to this end we integrate over a volume encompassing the whole dielectric body. In doing this, we can use Gauss's theorem to convert the $\nabla_j(\mathcal{P}_j\mathcal{E}_i^*+\mathcal{P}_j^*\mathcal{E}_i)$ volume integral into a surface integral enclosing the dielectric particle. However, the polarization is zero on this surface and so this term makes no contribution to the force and can be omitted. 

Finally, we average the fields over small volumes within the dielectric body to replace the them by the more familiar macroscopic fields \cite{Jackson1962}, writing 
\begin{equation}
\mathbfcal{P}=(\epsilon-1)\epsilon_0\mathbfcal{E},
\end{equation}
where $\epsilon$ is the complex relative permittivity of the dielectric body ($\epsilon=\epsilon'+i\epsilon''$). This then gives a total force of the form
\begin{align}
F_i&=\frac{1}{4}\int\left((\epsilon-1)\epsilon_0\mathcal{E}_j\nabla_i\mathcal{E}_j^* + (\epsilon^*-1)\epsilon_0\mathcal{E}_j^*\nabla_i\mathcal{E}_j\right) dV\nonumber\\
&=\frac{1}{4}\int\left((\epsilon'-1)\epsilon_0\nabla_i(\mathcal{E}_j\mathcal{E}_j^*)+2\epsilon''\epsilon_0(\mathcal{E}_j\mathcal{E}_j^*)\nabla_i\theta_j)\right)dV,
\end{align}
where $\theta_j=\arg(\mathcal{E}_j)$. The first contribution is the familiar dipole trapping force and the second is the recoil force due to absorption of the light \cite{Ashkinbook}. 

\section{Torque calculation}
We can readily extend the above calculation to angular momentum by replacing the linear momentum density and its flux by their angular analogues~\cite{Jackson1962}: $g_i\rightarrow(\mathbf{r}\times\mathbf{g})_i$, $T_{ij}\rightarrow \varepsilon_{ilm}r_lT_{mj}$. This procedure leads to a local conservation law for total angular momentum in which the force density in Eq. (\ref{eq:forcedensity}) is replaced by the total torque density
\begin{equation}\label{eq:torquedensity}
\tau_i=\varepsilon_{ijk}r_j\frac{1}{4}\left[\mathcal{P}_l\nabla_k\mathcal{E}_l^*+\mathcal{P}_l^*\nabla_k\mathcal{E}_l-\nabla_k(\mathcal{P}_l\mathcal{E}_k^*+\mathcal{P}_l^*\mathcal{E}_k)\right].
\end{equation}
We emphasise that, at this stage, it is essential to retain the third contribution as we have yet to perform the volume integral in order to determine the total torque. This component \textit{will} contribute to the total torque by virtue of the presence of the coordinate $r_j$ in the torque density. Evaluating the total torque, we find three contributions: 
\begin{align}\label{eq:22}
T_i
&=\frac{1}{4}\int dV\left({(\epsilon'-1)\epsilon_0(\mathbf{r}\times\mbox{\boldmath$\nabla$})_i(\mathcal{E}_j\mathcal{E}_j^*)}\right.\nonumber\\
&\left.\;\;\;\;\;\;\;\;+{2\epsilon''\epsilon_0\mathcal{E}_j\mathcal{E}_j^*(\mathbf{r}\times\mbox{\boldmath$\nabla$})_i\theta_j}+{2\epsilon''\epsilon_0 \Im (\mathbfcal{E}^*\times\mathbfcal{E})_i}\right).
\end{align}
The first of these  is the orbital part of the dipole trapping force, and the second term defines the transfer of orbital angular momentum to the material. The third, we identify as the transfer of spin angular momentum.
This is central result of this letter: the transfer of \textit{both} the spin and orbital components of optical angular momentum arise from considering the flux associated with  the optical angular momentum density $\mathbf{r}\times(\mathbf{E}\times\mathbf{H})/c^2$.
To make contact with experimental results \cite{ONeil2002}  we ask what is the contribution of the two absorptive processes on the motion of the trapped particle. The first term clearly introduces a force perpendicular to $\mathbf{r}$ and  imports an orbital angular momentum about the origin for a suitable beam. In particular for an orbital angular momentum beam with azimuthal phase dependence $\mathrm{e}^{il\phi}$, the net torque about the z-axis is 
\begin{align}
T_z^{\mathrm{orbital}}=\frac{l}{2}\int dV \epsilon''\epsilon_0\mathcal{E}_j\mathcal{E}_j^*, 
\end{align}
which is entirely natural. For the spin part, we note that the particle is trapped in such a way that its centre of mass does not move (there is no net force). It follows that the total spin contribution to the torque imparts a rotation of the trapped particle about its centre of mass: 
\begin{align}
T_z^{\mathrm{spin}}=\frac{1}{2}\int dV \epsilon''\epsilon_0 \Im (\mathbfcal{E}^*\times\mathbfcal{E})_z. 
\end{align}
For a paraxial beam we can write $\Im(\mathbfcal{E}^*\times\mathbfcal{E})_z=\mathcal{E}_j\mathcal{E}_j^*s$, where $s$ varies between $-1$ and $+1$ with its extreme values corresponding to the two possible circular polarizations. Hence for a particle trapped with its centre of mass at the centre of a tweezing beam we have 
\begin{align}
T_z=\frac{l+s}{2}\int dV \epsilon''\epsilon_0\mathcal{E}_j\mathcal{E}_j^*, 
\end{align}
and we can verify the equality of the two torques by cancelling them, as has been demonstrated experimentally \cite{Friese1996,Simpson1997}.

It is worthwhile pointing out that the total torque in Eq. (\ref{eq:22}) can be calculated directly from the continuity equation for the angular momentum $\mathbf{r}\times\mathbf{g}$ alongside the associated flux density \cite{Barnett2002}. We have instead chosen to start with the continuity equation for the \textit{linear} momentum density for two reasons, the first of which is simply to keep the calculation as straightforward as possible. Our second reason for beginning with the linear momentum and calculation of the force density has been to highlight that it is the total divergence term in Eq. (\ref{eq:torquedensity}) that is responsible for the spin contribution to the total torque. Due to the intrinsic nature of the spin angular momentum, this piece contributes zero total force, but calculation of the torque introduces the position vector $\mathbf{r}$ and hence a sense of direction within the system. 

The treatment above can be readily extended to magnetic media by means of a duality transformation of $\mathbf{E},\,\mathbf{H},\,\mathbf{P}$ and $\mathbf{M}$~\cite{Barnett2015a}, and has not been included explicitly here only for the purpose of brevity.  

\section{Conclusion}\label{sec:conclusion}
We have performed an \textit{ab initio} calculation of the separate spin and orbital contributions to the total electromagnetic torque, and have shown explicitly that \textit{both} of these pieces can be derived from the angular momentum flux  associated with the density  $\mathbf{r}\times(\mathbf{E}\times\mathbf{H})/c^2$. Crucially, the forms of the resulting torques agree with experimental evidence. The analysis presented has sought to close the question of whether the spin part  of the total angular momentum is contained in the angular momentum density defined by the cross product of the position vector with the linear momentum density: we have demonstrated that it is, by explicitly showing that both the spin and orbital torques can be derived from this piece when the flux density associated with this angular momentum is considered. This is done through direct calculation of the forces and torques exerted  on a piece of dielectric by  way of the momentum continuity equation. Finally, our analysis has emphasised the important point that, in general, both the angular momentum density and the flux density are physically significant, and, moreover, for monochromatic fields it is the angular momentum flux that is usually physically significant. 
\\ 

This work was supported by The Royal Society under Grant Numbers RF/ERE/231172 and RP150122.


\begin{thebibliography}{99}


 \bibitem{Allen1992} ALLEN L., BEIJERSBERGEN M. W., SPREEUW R. J. C. and WOERDMAN J. P., \textit{Phys. Rev. A} \textbf{45} {(1992)} {8185}.

\bibitem{OAMbook} {ALLEN L., BARNETT S. M. and PADGETT M. J.}, \textit{Optical Angular Momentum}, IOP Publishing, London {(2003)}.
  
   \bibitem{ONeil2002} {O'NEIL A. T., MACVICAR I., ALLEN L.,  and PADJETT M. J.}, \textit{Phys. Rev. Lett} \textbf{88} {(2002)} {4}.
  
    \bibitem{Friese1996} {FRIESE M. E. J., ENGER J., RUBINSZTEIN-DUNLOP H., and HECKENBERG N. R.}, \textit{Phys. Rev. A}{ \textbf{54}} {(1996)} {1593}.
  
  
  \bibitem{Simpson1997} {SIMPSON N. B., DHOLAKIA K., ALLEN L., and PADGETT M. J.},
 \textit{Opt. Lett.} {\textbf{22}} {(1997)} {52}


\bibitem{Jackson1962} {JACKSON J. D.},
 \textit{Classical Electrodynamics, second edition}, {Wiley, New York} {(1962)}
 
  
   \bibitem{Darwin1932} {DARWIN C. G.},
\textit{Proc. Roy. Soc. London, Series A} \textbf{136} {(1932)} {36}.
  
    \bibitem{Khrapko2001} {KHRAPKO R. I.}, \textit{Am. J. Phys.} \textbf{69} ({2001}) {405}.
 
  
  \bibitem{Simmons1970} {SIMMONS J. W., GUTTMANN M.}, \textit{States, Waves and Photons: A Modern Introduction to Light, First edition}, {Addison-Wesley, Massachusetts} {(1970)}
 
  
   \bibitem{Allen2002} {ALLEN L. and PADGETT M. J.},
  \textit{Am. J. Phys.} \textbf{70} {(2002)} {567}.
  
   \bibitem{Zambrini2005} {ZAMBRINI R. and BARNETT S. M.},
  \textit{J. Mod. Opt} \textbf{52} {(2005)} {1045}.
  
   \bibitem{Feynman1965} {FEYNMAN R. P., LEIGHTON R. B., SANDS M.}, \textit{The Feynman Lectures on Physics, Volume 3}, {Addison Wesley, London} {(1965)}.
  
  
   \bibitem{Barnett2002} {BARNETT S. M.},
  \textit{J. Opt. B: Quantum Semiclass. Opt.} \textbf{4} {(2002)} {7}.
  
  
   \bibitem{Marsden} {CRICHTON J. H. and MARSTON L.},
  \textit{Elec. J. of Differential Equations} \textbf{4} {(2000)} {37}.  
  
  
   \bibitem{Strasser2022} {STRASSER F., BARNETT S. M., RITSCH-MARTE M. and THALHAMMER G.},
  \textit{Phys. Rev. Lett.} \textbf{128} ({2002}) {213604}.
  
  
   \bibitem{Phillips} {PHILLIPS W. B}, \textit{Private communication}.
  
  
\bibitem{Lorentzbook} {LORENTZ H. A.},
  \textit{The Theory of Electrons}, {B. G. Teubner, Leipzig} (1909). 
  
  
     \bibitem{Barnett2006a} {BARNETT S. M. and LOUDON R.},
  \textit{J. PHYS. B: Atomic, Molecular and Optical Physics} \textbf{39} {(2006)} {S671}.
  
  
   \bibitem{Ashkinbook} {ASHKIN A.}, 
  \textit{Optical Trapping and Manipulation of Neutral Particles Using Lasers: A Reprint Volume with Commentaries}, {World Scientific Publishing Co. Pte. Ltd.} (2006)
  
    \bibitem{Barnett2015a} {BARNETT S. M. and LOUDON R.}, 
  \textit{New J. Phys.} \textbf{17} {(2015)} {063027}.

\end{thebibliography}
\end{document}